\documentclass[graybox,natbib]{SNmult}

\usepackage{type1cm}        
\usepackage{makeidx}         
\usepackage{graphicx}        
\usepackage{multicol}        
\usepackage[bottom]{footmisc}

\usepackage{amsmath}         
\usepackage{bm}              
\usepackage{url}             
\usepackage{hyperref}

\usepackage{newtxtext}       %
\usepackage[varvw]{newtxmath}       

\makeindex             

\begin{document}
\title*{3D mechano-geometric multicellular model of apical stem cell-driven plant morphogenesis}
\titlerunning{3D mechano-geometric model of plant morphogenesis}
\author{Naoya Kamamoto and Koichi Fujimoto}
\authorrunning{N. Kamamoto and K. Fujimoto}
\institute{Naoya Kamamoto \at Program of Mathematical and Life Science, Hiroshima University, 1-3-1 Kagamiyama, Higashi-Hiroshima City, Hiroshima 739-8526, Japan
\and Koichi Fujimoto \at Program of Mathematical and Life Science, Hiroshima University, 1-3-1 Kagamiyama, Higashi-Hiroshima City, Hiroshima 739-8526, Japan}
%
%
\maketitle

\abstract{
The orientation of cell division is a major determinant of three-dimensional plant morphogenesis. 
Whether and how a simple division orientation rule explains the establishment of symmetric body plans is a fundamental question. Testing such hypotheses is facilitated by 
a modeling framework that combines realistic three-dimensional cell mechanics, irreversible cell-wall growth, and a deformable tissue geometry. We recently introduced such a framework, a 3D mechano-geometric multicellular model of apical stem cell-driven morphogenesis. Here we document how the model is built from physiological and computational perspectives. We describe the triangulated thin-shell representation of cells, the treatment of turgor pressure, cell-wall elasticity and strain-driven wall growth, the cell-division algorithm together with its two pluggable division-rule implementations, and the remeshing operations that keep the triangulation well-conditioned as cells grow, divide, and deform. The aim of this paper is to make the present model accessible and customizable to experimental plant biologists.
\keywords{Plant morphogenesis, Apical stem cell, Cell division orientation, Mathematical model, Cell mechanics}}

\section{Introduction}
\label{sec:intro}

\begin{figure}
    \centering
    \includegraphics[width=1\linewidth]{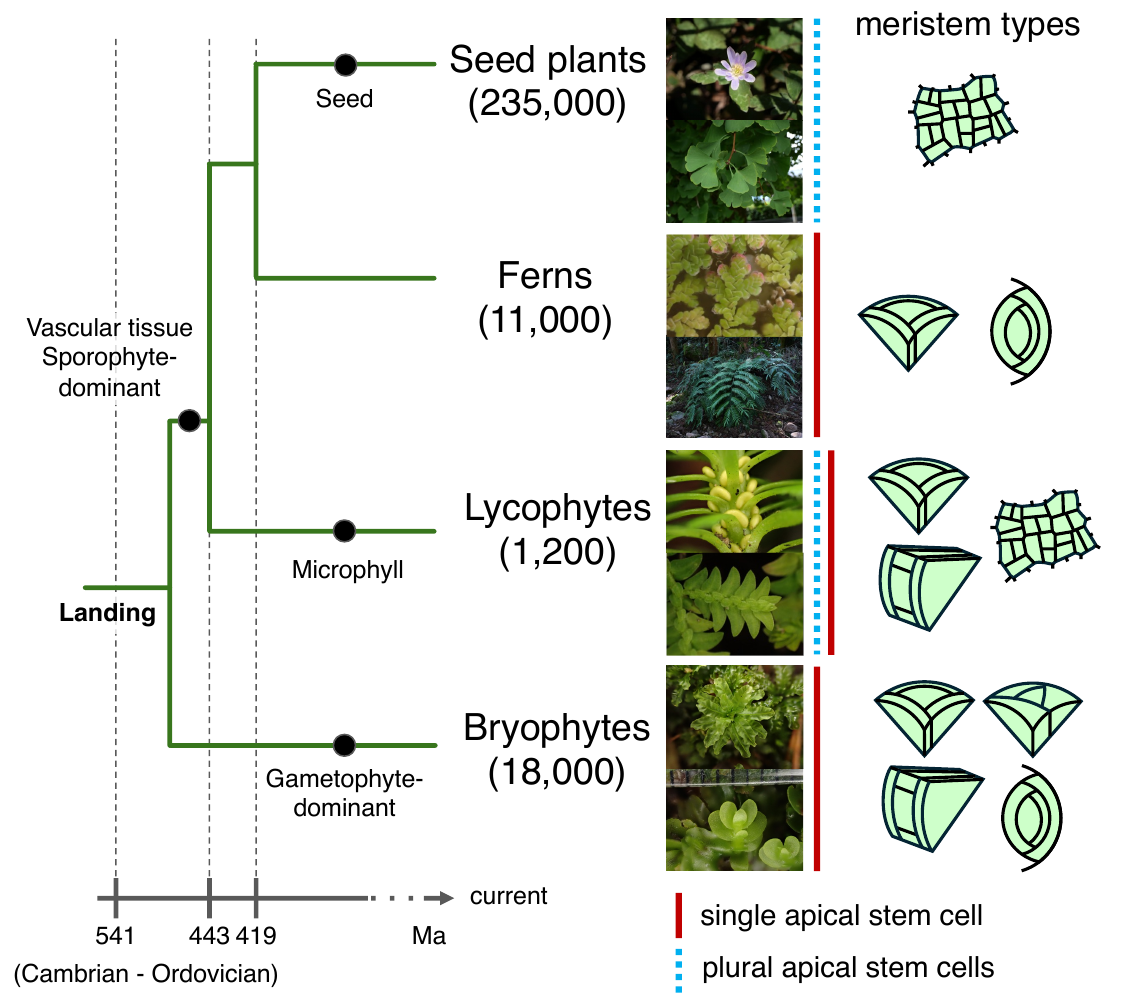}
    \caption{\textbf{Phylogeny of extant land plants and meristem types.} The phylogenetic tree is based on \citep{adl_revised_2012} and \citep{hasebe_morphology_2020}. The approximate number of extant species in each clade is shown in parentheses. Photographs from top to bottom show \textit{Anemone keiskeana}; \textit{Ginkgo biloba}; \textit{Azolla} sp.; \textit{Angiopteris lygodiifolia}; Sporangia of \textit{Huperzia serrata}; \textit{Selaginella heterostachis}; \textit{Plagiomnium maximoviczii}; \textit{Haplomitrium mnioides}.}
    \label{fig:Fig1}
\end{figure}

The precise orientation of cell division is fundamental for the 3D development of land plants \citep{smolarkiewicz_formative_2013}.
Because plant cells are encased in rigid cell walls and cannot rearrange, the position and orientation of every new division wall determines the architecture of the subsequent tissue.
This importance is most directly visible in basal land plants.
Bryophytes, ferns , and some lycophytes possess a single self-renewing apical stem cell (AC) at the growing tip (Fig. \ref{fig:Fig1}) \citep{gifford_morphology_1989, hasebe_morphology_2020}, and each AC division asymmetrically produces a self-renewing AC and a differentiated daughter cell (merophyte), which subsequently gives rise to a leaf and a portion of the stem \citep{douin_theorie_1925, gifford_jr_concept_1983, korn_apical_1993, harrison_local_2009}.
Successive AC divisions rotate around the AC by a species-specific or an organ-specific angle (typically 120$^\circ$ for the widely observed tetrahedral AC); thus, this rotational division alone is sufficient to specify the helically symmetric arrangement of lateral organs along the axis \citep{harrison_local_2009, veron_phyllotaxis_2021,kamamoto_rotation_2021}.
Consequently, in these clades the orientation of each AC division directly encodes the lateral organ arrangement (phyllotaxis).

In plants with multicellular apical meristems, namely the lycophytes, and the seed plants (Fig. 1), the link between AC division orientation and phyllotaxis is less direct, but division orientation still governs the supra-cellular geometry of the meristem.
In angiosperms, plural ACs reside at the center of the SAM, and phyllotaxis is patterned not by the AC division pattern but by the self-organizing transport of the plant hormone auxin cooperatively with PIN \citep{reinhardt_regulation_2003, jonsson_auxin-driven_2006, smith_plausible_2006}.
Even in this case, division orientation regulates the cell-adjacency topology of the meristem, namely the number of neighboring cells, the contact area between them, and the aspect ratio of each cell.
Consistent with this, a recent lineage-tracing study in \textit{Arabidopsis} has reported that each axillary branch traces back to a single progenitor cell in each of the three meristem layers \citep{xia_mapping_2026}, suggesting that the stem-cell division pattern remains more consequential in seed plants than has traditionally been appreciated.

Despite these repeated links between division orientation, meristem geometry, and organ arrangement, the molecular and physical actors that orient the division plane remain only partially understood.
Alternatively, two partially coincident hypotheses have been proposed to capture how the cell determines the division axis.
The least area (Errera's) rule posits that the division plane minimizes its area at a prescribed daughter-volume ratio \citep{errera_sur_1886, berthold_studien_1886, besson_universal_2011}, or passing through the mother cell center \citep{moukhtar_cell_2019}.
The maximal tension rule instead posits that the division plane is oriented parallel to the direction of maximal tension on the mother cell wall \citep{louveaux_mechanics_2013, hofler_mechanical_2024}.
Experimental evidence supports both rules across diverse tissues and species, and the two rules align closely under low stress anisotropy \citep{louveaux_mechanics_2013, guerin_forces_2016}.
Which rule operates in which context, and why, remain to be elucidated.

Mathematical and computational modeling has yielded fruitful results on this question, serving not only to generate experimentally testable predictions but also to reveal how particular cell-arrangement patterns stably arise from a given division orientation rule.
By specifying a candidate rule explicitly and iterating cell divisions on a simulated cell or tissue, one can ask whether the rule alone, given realistic geometry and mechanics, reproduces the observed cell arrangements. 
Besson and Dumais formalized the least area rule as a stochastic choice among local minima of the division-wall area, and identified a single ``inverse temperature'' parameter that remains constant across taxa while fitting division plane distributions from green algae to angiosperms \citep{besson_universal_2011}.
Subsequent studies extended the least area rule to fully three-dimensional mother cell geometries and applied it to the Arabidopsis embryo \citep{moukhtar_cell_2019} and other organs \citep{martinez_predicting_2018}, and moss Physcomitrium midlib cells \citep{ishikawa_gras_2023}. 
More recently, Couturier and colleagues elegantly showed that the shortest-geodesic version of the least area rule on a 2D conical meristem reproduces the three AC rotation angles (90$^\circ$, 120$^\circ$, and 180$^\circ$) observed across plants \citep{couturier_self-replicating_2025}, and Cammarata and colleagues, informed by live imaging of \textit{Physcomitrium patens}, combined the least area rule with a directed displacement of the division-plane centroid to reproduce rotational divisions in three dimensions \citep{cammarata_spiral_2026}.
In parallel, elaborate mechanical multicellular frameworks, which treat the cell wall as an elastic material under turgor and drive growth through wall yielding, have been used to test the maximal tension rule in the shoot apex \citep{louveaux_cell_2016} and cambium \citep{hofler_mechanical_2024}.

A missing element has been a single modeling framework in which both division rules can be tested on equal footing under plausible 3D cell mechanics and cell growth, and in a tissue geometry that is itself free to deform as cells grow and divide. 
We recently developed such a framework, a 3D mechano-geometric multicellular model of AC-driven morphogenesis, and used it to compare the least area and the maximal tension rules \citep{kamamoto_fujimoto_2026}.
Here, rather than revisiting the biological conclusions of that study, we document how the model is built.
We describe the triangulated thin-shell representation of cells, the treatment of turgor pressure, cell-wall elasticity, and cell-wall growth, the cell-division algorithm together with the two division-rule implementations it supports, and the remeshing operations that keep the triangulation well-conditioned as cells grow, divide, and deform.
Our aim is to make this class of 3D mechano-geometric model accessible and modifiable to experimental plant biologists, and thereby to lower the barrier for adopting and critically evaluating modeling-driven approaches in this field.

\section{Methods}
\label{sec:model}

\subsection{Overview}
\begin{figure}
    \centering
    \includegraphics[width=1\linewidth]{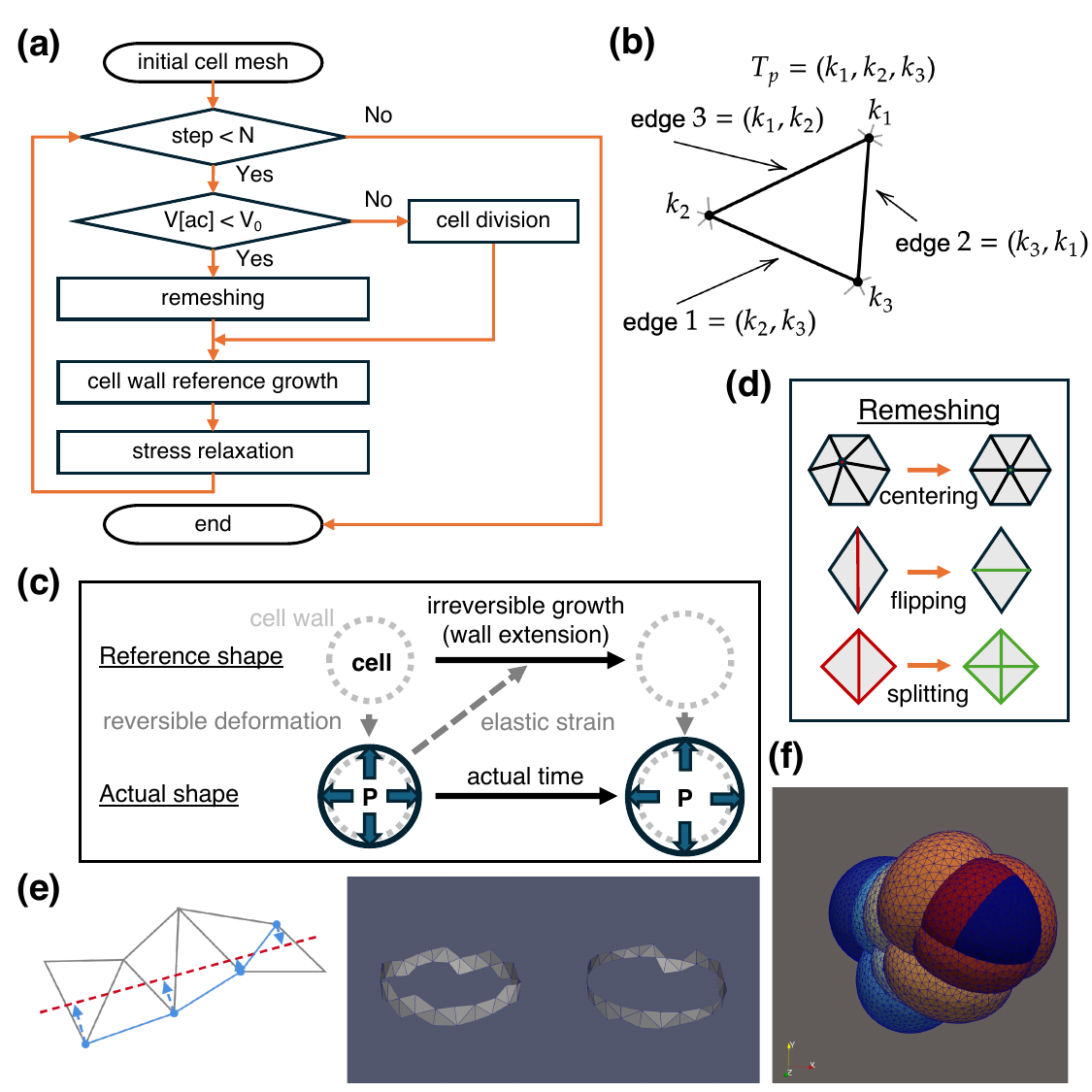}
    \caption{\textbf{Overview of the model implementation}.
    (a) Simulation workflow. 
    (b) Triangle element and related variables. 
    (c) Deformation and growth. 
    (d) Remeshing operations. 
    (e) Construction of the cortical division zone. Left: projection of closed-loop vertices (blue circles) onto the upcoming division plane (red dashed line). Center and right: face elements around the CDZ before and after projection and vertex centerings.
    (f) Helically symmetric cell arrangement obtained after 12 least area divisions of the AC. Parameters are the same as in \citep{kamamoto_fujimoto_2026}. 
    }
    \label{fig:Fig2}
\end{figure}

We represent each cell as a closed, triangulated thin shell whose triangular elements along a common wall are shared with the neighboring cell.
Cell adhesion is thus imposed by mesh construction rather than by an explicit contact model. 
Plant development is modeled as the interplay of three mechanical processes: cell-wall deformation driven by a constant turgor pressure, slow irreversible cell-wall growth, and division of the AC. 
Each simulation time step consists of mechanical relaxation, local remeshing, growth of the reference edge lengths, and a cell division check  (Fig. \ref{fig:Fig2}a).
The implementation is in Julia \citep{bezanson_julia_2017} and results are visualized using ParaView \citep{ahrens_paraview_2005}.
The simulation code is available at \url{https://github.com/NaoyaKamamoto/Plant-3d-multicellular-model}. 

\subsection{Mesh representation and data structure}
\label{subsec:mesh}
Because the cell wall is a thin curved surface and its shape is the object of the simulation, we discretize it as a triangular mesh. 
This representation allows mechanical deformation and cell wall growth to be implemented as local updates of vertices, edges, and faces. 

The cell-wall geometry is stored as (i) a list of vertex positions $\{\bm{x}_k\}$, (ii) a list of triangular faces $\{T_p\}$, each specified by three vertex indices, and (iii) a list of edges, each a pair of vertex indices (Fig. \ref{fig:Fig2}b).
Each edge stores a reference length $L_j$; this reference quantity defines the local stress-free configuration against which the actual, deformed length $l_j$ is compared (Fig. \ref{fig:Fig2}c).

\subsection{Cell wall as a hyperelastic thin shell}
\label{subsec:cellwall}
Plant cell walls are thin, stiff layers of polysaccharide that bear the mechanical load exerted by turgor pressure (Fig. \ref{fig:Fig2}c).
For the morphogenetic processes considered here, we assume that in-plane stretching is the dominant mode of deformation, whereas bending and through-thickness deformation are neglected. 
We therefore model each wall as an elastic thin shell and use a hyperelastic constitutive law to describe finite in-plane strains during cell growth.
This continuum mechanical model is then discretized using the TRBS formulation introduced below, which expresses the strain energy of each triangle as a closed-form function of its edge lengths and provides analytical gradients for energy minimization \citep{delingette_biquadratic_2008}.

The cell wall is treated as a thin shell of St.~Venant-Kirchhoff hyperelastic material, whose continuum strain energy is
\begin{align}
W_\Omega = \int_\Omega \left( \frac{\lambda}{2} (\mathrm{tr}\,\bm{E})^2 + \mu\, \bm{E}:\bm{E} \right) d\Omega,
\end{align}
where $\bm{E}$ is the Green-Lagrange strain tensor and $\lambda$, $\mu$ are the Lam\'e constants, given in terms of Young's modulus $Y$ and Poisson's ratio $\sigma$ by $\lambda = Y\sigma/(1-\sigma^2)$ and $\mu = Y/\{2(1+\sigma)\}$. 
We discretize this shell using the triangular biquadratic spring (TRBS) model \citep{delingette_biquadratic_2008}, which has been applied previously to plant multicellular modeling \citep{bozorg_stress_2014, bonfanti_stiffness_2023}.
The TRBS formulation exactly reproduces the continuum energy on each triangle as a biquadratic function of the squared edge-length deviations from the reference configuration \citep{delingette_biquadratic_2008}:
\begin{align}
W_\mathrm{TRBS}(T_p) = \sum_{i=1}^3 \frac{k_i^{T_p}}{4} \left(l_i^2 - L_i^2\right)^2 + \sum_{i \neq j} \frac{c_{ij}^{T_p}}{2}\left(l_i^2 - L_i^2\right)\left(l_j^2 - L_j^2\right),
\end{align}
where $l_i$ and $L_i$ are the actual and reference lengths of the $i$-th edge of triangle $T_p$.
The effective spring constants are given by
\begin{align}
k_i^{T_p} = \frac{2\cot^2\alpha_i (\lambda+\mu) + \mu}{16 A_p}, \qquad c_{ij}^{T_p} = \frac{2\cot\alpha_i\cot\alpha_j (\lambda+\mu) - \mu}{16 A_p},
\end{align}
where $A_p$ is the area of the reference triangle and $\alpha_i$ is the interior angle opposite the $i$-th edge in the reference configuration.
These constants are precomputed for every triangle and updated whenever the reference configuration changes.
The total elastic energy of the cell sheet is $W_\mathrm{el} = \sum_p W_\mathrm{TRBS}(T_p)$, the sum running over every triangle, including shared walls.
Because a shared wall is stored as a single triangular element but represents the apposing wall layers of two neighboring cells, we double its elastic stiffness (equivalently, its spring constants $k_i^{T_p}$ and $c_{ij}^{T_p}$) relative to a free-surface triangle of the same geometry.

In practical terms, each edge behaves as a nonlinear spring: the reference length $L_i$ defines the local stress-free shape that is modified during growth, while $l_i$ is the actual length after mechanical relaxation. 

\subsection{Turgor pressure}
\label{subsec:turgor}
Turgor is the hydrostatic pressure exerted by the  cytoplasm on the cell wall, and it is the primary driver of plant cell expansion.
We treat it as piecewise constant (one value per cell, constant in time) because its redistribution by water fluxes is fast relative to the growth timescale and is not the focus of this model.

For each triangular face $T_p$, the signed pressure load is computed from the pressure difference $\Delta P_i$ across the face. 
The corresponding force is distributed equally among the three vertices of the triangle. 
For vertex $k$, this force is
\begin{align}
\bm{f}_k^{(p)} = \frac{\Delta P_i}{6}\, \bm{x}_{k'} \times \bm{x}_{k''},
\end{align}
where $(k, k', k'')$ are the three vertex indices of the triangle in the counter-clockwise manner. 
The total pressure force on a vertex is obtained by summing this contribution over all connected triangles.
Because $\Delta P_i$ is defined as a pressure difference, a shared wall between two cells with equal turgor receives no net pressure force.

\subsection{Stress relaxation}
\label{subsec:relaxation}
Plant cell walls relax mechanically on a timescale much shorter than that of appreciable growth.
We therefore assume quasi-static mechanics: after each growth update, the tissue is brought to mechanical equilibrium before the next growth step is applied. 
This equilibrium is defined as the configuration in which the total force on every free vertex vanishes, and is computed by minimizing the total potential energy
\begin{align}
U(\{\bm{x}_k\}) = \sum_p W_\mathrm{TRBS}(T_p) - \sum_c P_c V_c(\{\bm{x}_k\}),
\end{align}
with respect to the vertex coordinates $\{\bm{x}_k\}$, where $V_c$ is the volume of cell $c$.
The minimization is performed using the L-BFGS method \citep{liu_limited_1989}. 

\subsection{Cell wall growth}
\label{subsec:growth}
Growth of a plant cell wall is not a simple elastic stretch: new wall material is continuously deposited and existing material is remodeled, so that sustained tension irreversibly relaxes into permanent elongation. 
We capture this by letting the reference length $L_j$ of each edge grow when the edge is stretched beyond a yield strain, while the current length $l_j$ is determined by mechanical equilibrium. 
The formulation is a per-edge version of the classical Lockhart-Ortega rheology adapted to a triangulated surface. 

Cell-wall growth is modeled as the irreversible elongation of each reference edge length $L_j$, following the strain-based formulation of previous studies \citep{bassel_mechanical_2014, bonfanti_stiffness_2023},
\begin{align}
L_j^\mathrm{new} = L_j + \Delta t\, \Phi_j \left[ \frac{l_j}{L_j} - 1 - \varepsilon_y \right]_+ L_j,
\label{eq:growth}
\end{align}
where $\Phi_j$ is the extensibility of edge $j$, $\Delta t$ the growth time step, $\varepsilon_y$ the yield strain, and $[\,\cdot\,]_+$ the ramp function (zero for non-positive arguments, identity otherwise).
Each edge grows only when its elastic strain exceeds $\varepsilon_y$; otherwise, its reference length remains unchanged.

\subsection{Remeshing}
\label{subsec:remesh}
The mesh is a computational device rather than a biological entity, and the simulation should therefore be insensitive to its exact layout. 
As cells grow and divide, however, triangles can become highly irregular, which reduces the accuracy of stress relaxation and may cause artifacts during later divisions. 
To maintain mesh quality, we periodically apply three local remeshing operations: vertex centering, edge splitting, and edge flipping (Fig. \ref{fig:Fig2}d) \citep{okuda_continuum_2022, khan_surface_2022}. 

To avoid introducing artificial changes in the mechanical state, the reference lengths are updated so that the local strain is preserved as closely as possible. 
When an edge is modified by remeshing operations, its new reference length $L'$ is computed from the current length $l'$ using the strain-preserving rule of \citet{bozorg_continuous_2016}:
\begin{align}
L' = \left[ \sum_i (1 - 2\lambda_i)\, |\bm{t}\cdot\bm{s}_i|^2 \right]^{1/2} l',
\label{eq:bozorg}
\end{align}
where $\lambda_i$ and $\bm{s}_i$ are the eigenvalues and eigenvectors of the mean Euler-Almansi strain tensor of the incident triangles and $\bm{t}$ is the unit tangent vector of the edge.
This update keeps the mechanical state before and after remeshing approximately consistent.
\newline

\noindent\textbf{Vertex centering:}
For every vertex $\bm{x}$ not on the cortical division zone (see below), we compute the area-weighted centroid of its incident triangles,
\begin{align}
\bm{c} = \frac{\sum_i a_i \bm{c}_i}{\sum_i a_i},
\end{align}
where $a_i$ and $\bm{c}_i$ are the area and centroid of the $i$-th incident triangle, respectively. 
We also compute the area-weighted mean outward normal,
\begin{align}
\bm{n} = \frac{\sum_i a_i \bm{n}_i}{|\sum_i a_i\bm{n}_i|},
\end{align}
where $\bm{n}_i$ is the outward unit normal of the $i$-th incident triangle. 
The vertex is then relocated along this mean-normal direction to 
$\bm{x}_\mathrm{new} = \bm{c} + h\bm{n}$,
where the scalar $h$ is chosen so that the surrounding cell volume is unchanged.
Specifically, $h$ is found by solving
$(V(\bm{x}_\mathrm{new}) - V_0)^2 = 0$,
where $V(\bm{x}_\mathrm{new})$
and $V_0$ are the cell volume after and before the centering, respectively.
The operation is triggered only when the expected displacement exceeds $0.1 \times \min \{L_j^{(0)}\}$, where $\{L_j^{(0)}\}$ are the reference edge lengths at the initial of simulation.
The new reference lengths of the incident edges are updated by Eq.~\eqref{eq:bozorg}.
For vertices on the cortical division zone (during division, below), a variant that conserves surface area rather than volume is used, and the new reference length of each incident edge is set to $0.99 l'$.
\newline

\noindent\textbf{Edge splitting:}
An edge is considered for splitting when both the total area of its incident triangles exceeds a prescribed threshold and its length exceeds
$l_\mathrm{max} = 1.1 \times \max \{L_j^{(0)}\}$.
For a manifold edge shared by two triangles, a new vertex is inserted at the edge midpoint, and each adjacent triangle is divided into two, producing four triangles. 
The same midpoint-splitting operation is also applied to non-manifold edges at tricellular or cortical junctions, with all incident triangles updated accordingly. 
For manifold edges, the reference lengths of the newly created edges are computed using Eq.~\eqref{eq:bozorg}.
For non-manifold edges, where a unique mean strain tensor is not well defined, the new reference length is set to $0.99l'$. 
To avoid the generation of narrow triangles, a split is accepted only if all resulting triangles satisfy the regularity condition
$(a+b-c)/(a+b+c) \geq r_\mathrm{min} = 0.1$,
where $a \leq b \leq c$  are the sorted edge lengths of each resulting triangle.
To avoid cascading stress perturbations, an edge is not split if any of its incident triangles has already been split in the same time step.
\newline

\noindent\textbf{Edge flipping:}
A poorly shaped pair of triangles is improved by replacing their shared diagonal with the opposite diagonal. 
Flipping is restricted to manifold edges whose graph-theoretic degree is greater than three, and is performed only when the change in diagonal edge length after flipping exceeds $0.1 \times \min \{L_j^{(0)}\}$.
This avoids unnecessary flips that have little effect on mesh quality.
We also require the four angles adjacent to the new diagonal in the post-flip configuration to be smaller than $5\pi/12 =75^\circ$.
This condition prevents the creation of locally concave configurations that can lead to inverted or overlapping triangles.
The reference length of the new diagonal is computed using Eq.~\eqref{eq:bozorg}.

\subsection{Cell division}
\label{subsec:division}
A cell division in this model proceeds in three stages: deciding the position and the orientation of new cell wall, detecting and constructing the cortical division zone where the new wall attach, and constructing division wall as a mesh.
The mesh-insertion stage carries most of the implementation complexity, because all subsequent operations assume that the mesh remains a clean, a closed triangulated surface.
We describe these three stages in turn, axis selection by the least area or the maximal-tension rule, construction of the cortical division zone, and construction of the new division wall.
\newline

\noindent\textbf{Axis selection by the least area rule:}
The least-area rule has no closed-form solution for a general triangulated mother cell. 
We therefore approximate the minimum by evaluating the area of the candidate division wall over a finely sampled set of orientations and selecting the orientation with the least area. 
We parametrize the unit normal of the candidate division plane in spherical coordinate as:
\begin{align}
\bm{d}(\theta,\phi) = (\sin\theta\cos\phi,\, \sin\theta\sin\phi,\, \cos\theta),
\end{align}
and sample $(\theta,\phi)$ on a $180 \times 180$ uniform grid covering $\theta, \phi \in [0,\pi)$. 
This samples one hemisphere, which is sufficient because $\bm{d}$ and $-\bm{d}$ define the same division plane, and gives an angular resolution of 1$^\circ$. 
For each candidate orientation, we compute the area of the polygon formed by the intersection between the candidate plane and the triangulated surface of the mother cell. 
Specifically, we identify the surface triangles crossed by the plane, compute the intersection segment on each crossed triangle, assemble these segments into a planar polygon, and calculate its area. 
Finally, the axis $\bm{d}^*$ that gives the least intersection area over the sampled orientations is selected.
\newline

\noindent\textbf{Axis selection by the maximal-tension rule:}
The maximal tension rule uses the same grid of candidate axes as the least area rule, but evaluates each axis from the stress state of the mother cell rather than from wall area.
It selects the axis that best aligns the new division wall with the local directions of maximum tension on the free surface. 
On each triangular element $T_p$ of the mother cell's free surface, we compute the second Piola-Kirchhoff stress tensor from the current TRBS state.
Let $s_1^{(p)} \geq s_2^{(p)}$ be its principal stresses, and let $\bm{p}_p$ be the minor principal direction, defined as the unit in-plane vector perpendicular to the direction of maximum tension.
Thus, $\bm{p}_p$ is perpendicular to both the face normal $\bm{n}_p$ and the local maximum-tension direction.

For each candidate axis $\bm{d}$ sampled on the same $180\times 180$ grid, we evaluate
\begin{align}
S(\bm{d}) =
\sum_p w_p\, |\bm{d}\cdot\bm{p}_p|,
\qquad
w_p = a_p \left(s_1^{(p)} - s_2^{(p)}\right),
\label{eq:maxtension}
\end{align}
where the sum runs over the free-surface triangles and $a_p$ is the area of $T_p$.
The weight $w_p$ combines triangle area and stress anisotropy, so that large, highly anisotropic, tense regions contribute most strongly, whereas isotropically stressed regions contribute little.
The selected axis $\bm{d}^*$ is the one that maximizes $S(\bm{d})$.

Because $\bm{p}_p$ is perpendicular to the local maximum-tension direction in the tangent plane, maximizing $|\bm{d}\cdot\bm{p}_p|$ aligns the division-wall normal with the local minimum-tension direction.
Equivalently, the division wall itself tends to align with the local maximum-tension direction.
\newline

During candidate-axis selection, we exclude division planes that would pass through an existing four-cell junction. Such cases are rare, but they would require inserting a new division wall through a high-valence junction, which is not supported by the current mesh-surgery procedure. 
\newline

\noindent\textbf{Construction of the cortical division zone:}
The cortical division zone (CDZ) is the closed curve on the mother cell surface along which the new division wall will attach to the pre-existing wall. 
A naive implementation would cut every triangle intersected by the division plane, but this would create many highly elongated triangles and destabilize subsequent mechanical relaxation.
Alternatively, the CDZ is constructed by finding the shortest closed loop along existing mesh edges and projecting it onto the division plane, thereby minimizing perturbations to the surrounding triangulation (Fig. \ref{fig:Fig2}e). 
The projection is separated into several incremental steps.
Between projection steps, volume-conserving centerings are applied to the surrounding mother cell mesh to minimize perturbations to the cell shape and mechanical state. 
\newline

\noindent\textbf{Construction of the new division wall:}
Once the CDZ has been fixed and flattened into the division plane, a new triangulated wall is constructed to fill its interior. 
The new wall must satisfy two competing requirements: its vertices should be distributed uniformly enough to avoid thin triangles, but they must also remain compatible with the pre-existing CDZ vertices on the boundary. 
To achieve this, we use the bubble-mesh method \citet{shimada_bubble_1995}: fictitious disks (``bubbles'') of prescribed radius $r$ are initialized inside the CDZ and subjected to repulsive pairwise forces, and their equilibrium configuration provides a quasi-uniform set of vertex positions.
The pairwise force between bubbles $i$ and $j$ of radii $r_i$ and $r_j$ at separation $\bm{r}_{ij}$, with $\ell = \|\bm{r}_{ij}\|/(r_i+r_j)$, is taken as
\begin{align}
\bm{F}_{ij} = -w(\ell)\, \hat{\bm{r}}_{ij},\qquad
w(\ell) = \begin{cases} a(1-\ell) & \ell < 1,\\ b(1-\ell) & 1 \leq \ell < 2,\\ 0 & \ell \geq 2,\end{cases}
\end{align}
so that overlapping bubbles repel strongly (positive coefficient $a$) and non-overlapping but close bubbles attract weakly toward the preferred packing separation (positive coefficient $b < a$ in magnitude, taken so that the force remains repulsive but weaker).
The initial bubble radii are chosen to produce an areal density matching the average edge length on the adjacent mother-cell wall, and a ``modify-radii'' step adjusts the radii after a preliminary equilibration so that the final packing is homogeneous.

The bubble centers and the CDZ vertices are then triangulated in the plane by a constrained Delaunay triangulation, treating the CDZ edges as required boundaries.
The resulting set of triangles becomes the new shared wall between the two daughter cells.
The reference edge lengths of the newly created edges are initialized to $0.99\, l'$ (where $l'$ is the actual length in the new shared wall) to avoid numerical instability. 

\subsection{Default parameter values}
\label{subsec:params}
The simulations use Young's modulus $Y$ = 100 MPa, Poisson's ratio $\sigma = 0.2$, and turgor pressure $p_0 = 0.2$ MPa, following the previous study \citep{bonfanti_stiffness_2023}. 
The yield strain is set to $\varepsilon_y = 0$, so that stretched edges grow, whereas compressed edges do not shorten. 
The wall extensibility $\Phi$ is chosen so that one AC cell cycle corresponds to approximately 100 growth steps, keeping the strain increment per step small relative to the quasi-static relaxation timescale.
Cell division is triggered when the AC volume reaches twice its initial value.

\section{Acknowledgements}
We thank Masaki Shimamura (Hiroshima University, Japan), Satoru Tsugawa (Akita Pref. University, Japan), and Minako Ueda (Tohoku University, Japan) for valuable discussions and suggestions. This work is supported by the Japan Science and Technology Agency (CREST [JPMJCR2121 to N.K. and K.F.]) and the Japan Society for the Promotion of Science (26H00457 to K.F.).

\section{Competing Interest Statement}
The authors declare no competing interests.

%
%

\bibliographystyle{spbasic}
\bibliography{references}

\end{document}